# Measuring nuclear reaction cross sections to extract information on neutrinoless double beta decay

**M. Cavallaro[1*], F. Cappuzzello[1,2], C. Agodi[1], L. Acosta[3], N. Auerbach[4], J. Bellone[1,2], R. Bijker[3], D. Bonanno[5], D. Bongiovanni[1], T. Borello-Lewin[6], I. Boztosun[7], V. Branchina[2,5], M. P. Bussa[8,9], S. Calabrese[1,2], L. Calabretta[1], A. Calanna[1], D. Calvo[8], D. Carbone[1], E.R. Chávez Lomelí[3], A. Coban[7], M. Colonna[1], G. D'Agostino[1,2], G. De Geronimo[10], F. Delaunay[8,11], N. Deshmukh[1], P.N. de Faria[12], C. Ferraresi[8], J.L. Ferreira[1,12], P. Finocchiaro[1], M. Fisichella[8], A. Foti[2,5], G. Gallo[1,2], U. Garcia[13], G. Giraudo[8], V. Greco[1,2], A. Hacisalihoglu[1], J. Kotila[14], F. Iazzi[8,15], R. Introzzi[8,15], G. Lanzalone[1,16], A. Lavagno[8,9], F. La Via[1,17], J.A. Lay[18], H. Lenske[19], R. Linares[12], G. Litrico[1], F. Longhitano[5], D. Lo Presti[2,5], J. Lubian[12], N. Medina[6], D. R. Mendes[12], A. Muoio[1], J.R.B. Oliveira[6], A. Pakou[20], L. Pandola[1], H. Petrascu[21], F. Pinna[8,15], S. Reito[5], D. Rifuggiato[1], M.R.D. Rodrigues[6], A. D. Russo[1], G. Russo[2,5], G. Santagati[1], E. Santopinto[13], O. Sgouros[20], S.O. Solakcı[7], G. Souliotis[22], V. Soukeras[20], A. Spatafora[1,2], D. Torresi[1], S. Tudisco[1], R.I.M. Vsevolodovna[13], R.J. Wheadon[8], A. Yildirin[7], V.A.B. Zagatto[1,12]**

[1]Istituto Nazionale di Fisica Nucleare, Laboratori Nazionali del Sud, Catania, Italy
[2]Dipartimento di Fisica e Astronomia, Università di Catania, Italy
[3]Universidad Nacional Autónoma de México
[4]School of Physics and Astronomy Tel Aviv University, Israel
[5]Istituto Nazionale di Fisica Nucleare, Sezione di Catania,
[6]Universidade de Sao Paulo, Brazil
[7]Akdeniz University, Antalya, Turkey
[8]Istituto Nazionale di Fisica Nucleare, Sezione di Torino
[9]Università di Torino, Torino, Italy
[10]Stony Brook University, USA
[11]Université de Caen and Laboratoire de Physique Corpusculaire de Caen, France
[12]Universidade Federal Fluminense, Niteroi, Brazil
[13]Istituto Nazionale di Fisica Nucleare, Sezione di Genova
[14]University of Jyväskylä, Jyväskylä, Finland
[15]Politecnico di Torino, Italy
[16]Università degli Studi di Enna "Kore", Enna, Italy
[17]CNR-IMM, Sezione di Catania, Italy
[18]University of Seville, Spain
[19]University of Giessen, Germany
[20]University of Ioannina, Ioannina, Greece
[21]IFIN-HH, Romania
[22]University of Athens and HINP, Athens, Greece

*manuela.cavallaro@lns.infn.it

**Abstract**. Neutrinoless double beta decay (0νββ) is considered the best potential resource to access the absolute neutrino mass scale. Moreover, if observed, it will signal that neutrinos are their own anti-particles (Majorana particles). Presently, this physics case is one of the most important research "beyond Standard Model" and might guide the way towards a Grand Unified Theory of fundamental interactions.
Since the 0νββ decay process involves nuclei, its analysis necessarily implies nuclear structure issues. In the NURE project, supported by a Starting Grant of the European Research Council (ERC), nuclear reactions of double charge-exchange (DCE) are used as a tool to extract information on the 0νββ Nuclear Matrix Elements. In DCE reactions and ββ decay indeed the initial and final nuclear states are the same and the transition operators have similar structure. Thus the measurement of the DCE absolute cross-sections can give crucial information on ββ matrix elements. In a wider view, the NUMEN international collaboration plans a major upgrade of the INFN-LNS facilities in the next years in order to increase the experimental production of nuclei of at least two orders of magnitude, thus making feasible a systematic study of all the cases of interest as candidates for 0νββ.

## 1. The physics case

The potential observation of 0νββ decay, besides establishing the Majorana nature of neutrinos, has the potential to shed light on the absolute neutrino mass scale. To this purpose, it is critical that the associated Nuclear Matrix Elements (NME) are known with sufficient accuracy. The 0νββ decay rate can be expressed as a product of independent factors: the phase-space factors, the NME and a function of the masses of the neutrino species. Thus the knowledge of the NME can give information on the neutrino mass scale, if the 0νββ decay rate is measured. From a comparison of the results of different NME calculations, obtained within various nuclear structure frameworks [1–4], significant differences are found, even of a factor 2 or 3, which makes the present situation still not satisfactory. In addition, some assumption common to different competing calculations, like the unavoidable truncation of the many body wave-function, could cause overall systematic uncertainties.

The use of heavy-ion double charge exchange reactions (HI-DCE) as a tool in order to access quantitative information relevant for 0νββ decay NME has been recently proposed [5, 6]. These reactions are characterized by the transfer of two charge units, leaving the mass number unchanged, and can proceed by a sequential nucleon transfer mechanism or by meson exchange. Despite the fact that 0νββ decay and HI-DCE reactions are mediated by different interactions, they present a number of similarities. Among that, the key aspects are that initial and final nuclear states are the same and the transition operators in both cases present a superposition of isospin, spin-isospin and rank-two tensor components with a relevant momentum (100 MeV/c or so) available.

## 2. The technique

In a pioneering experiment, performed at the INFN-LNS laboratory, we studied the DCE reaction $^{40}$Ca($^{18}$O,$^{18}$Ne)$^{40}$Ar at 270 MeV, with the aim to measure the cross section at zero degrees [5]. In the experiment an advantageous condition was set, using a beam of $^{18}$O and a double magic target as $^{40}$Ca and choosing the bombarding energy to mismatch the competing transfer reactions leading to the same outgoing channel. The key tools in the experiment were the high resolution Superconducting Cyclotron beams and the MAGNEX spectrometer [7-9], a modern high resolution and large acceptance magnetic system [10-13] characterized by high resolution in energy, mass and angle. The high-order solution of the equation of motion is the key feature of MAGNEX, which guarantees the above mentioned performances and its relevance in the research of nuclear structure and reaction mechanisms [14-19]. In the "pilot experiment" we have shown that high resolution and statistically significant experimental data can be measured for DCE processes and that precious information towards NME determination could be at our reach.

The NURE (NUclear REactions for neutrinoless double beta decay) project has been selected for receiving funding in the call Starting Grant 2016 of European Research Council (ERC). The project has a duration of 5 years and plans to carry out a campaign of experiments by using heavy-ion accelerated beams impinging on two targets of interest as candidate nuclei for the ββ decay. The DCE channel will be explored by the ($^{18}$O,$^{18}$Ne) reaction for the β+β+ direction in the target and the ($^{20}$Ne,$^{20}$O) for the β-β-. Moreover, the complete net involving the multi-step transfer processes characterized by the same initial and final nuclei will be also studied in the same experimental setup. In particular, the ($^{18}$O,$^{16}$O) and ($^{20}$Ne,$^{22}$Ne) two-neutron transfer, ($^{18}$O,$^{20}$Ne) and ($^{20}$Ne,$^{18}$O) two-proton transfer, ($^{18}$O,$^{18}$F) and ($^{20}$Ne,$^{20}$F) single charge-exchange will be measured.

## 3. The optimization of the experimental apparatus

During the first months of the project, tests on the facility have been done in order to find the best compromise between energy resolution and count rate. In particular, a work on the optimization of the gas section of the MAGNEX focal plane detector [20] have been performed. A new design for the multiplication region, based on six position sensitive proportional wires has been developed and installed.

A special care has been devoted on the target production technologies, in order to optimize uniformity and thickness of the target films. Energy loss and spreading calculations have been done to find the optimal compromise between target thickness and requested resolution in the final energy spectra. This is a crucial point since high energy resolution is necessary to separate the $0^+$ ground state (which is the transition of interest) from the excited states of the final populated nucleus. On the other hand, a relevant target thickness is important to increase the count rate at the detector, since the expected cross-sections are very low (of the order of hundreds of nb/sr).

## 4. The NUMEN program

The possibility to extract data-driven information on 0νββ nuclear matrix elements is an ambitious line of research which has been conceived and is developing in a broader context, characterized by a longer time scale and a larger collaboration: the NUMEN program [6, 21].

In the present experimental conditions, due to the limitation arising from the tiny cross-sections of the processes of interest, only very few systems can be measured within the 5-years ERC project. In order to systematically explore all the nuclei candidates for 0νββ, a beam intensity at least two orders of magnitude higher than the present must be achieved [22]. The challenge is to measure a rare nuclear transition under a very high rate of heavy ions produced by the beam-target interaction. As a consequence, major upgrades of the detector technologies (3D ion tracker, particle-identification wall, gamma-ray array, …) must be developed [23, 24]. Also the target technology must be upgraded, to avoid the damage of the thin films due to the high temperature involved [25, 26]. The front-end and readout electronics must take into account the high number of channels and the expected rate of detected events. Moreover, a deep and complete investigation of the theoretical aspects connecting nuclear reaction mechanisms and nuclear matrix elements must be carried out.

All of these issues and other aspects related to the high intensity beam physics are the crucial branches of the NUMEN program.

## Acknowledgments

This project has received funding from the European Research Council (ERC) under the European Union's Horizon 2020 research and innovation programme (grant agreement No 714625).

## References

[1] Vergados J D *et al.* 2012 *Rep. Progr. Phys.* **75** 106301.

[2] Vogel P 2012 *Journal of Physics G: Nuclear and Particle Physics* **39** 124002
[3] Engel J and Menendez J 2017 *Rep. Prog. Phys.* **80** 046301
[4] Barea J *et al.* 2015 *Phys. Rev. C* **91** 034304
[5] Cappuzzello F *et al.* 2015 *Eur. Phys. J. A* **51**:145
[6] Cappuzzello F *et al.* 2015 *Journal of Physics Conference Series* **630** 012018
[7] Cappuzzello F *et al.* 2016 *Eur. Phys. J. A* **52**: 167
[8] Cunsolo A *et al.* 2007 *Eur. Phys. J. A ST* **150** 343
[9] Cappuzzello F *et al.* 2014 *Nucl. Instr. And Meth. A* **763** 314
[10] Lazzaro A *et al.* 2008 *Nucl. Instr. and Methods A* **591** 394
[11] Lazzaro A *et al.* 2008 *Nucl. Instr. and Methods A* **585** 136
[12] Lazzaro A *et al.* 2007 *Nucl. Instr. and Methods A* **570** 192
[13] Lazzaro A *et al.* 2009 *Nucl. Instr. and Methods A* **602** 494
[14] Cappuzzello F *et al.* 2015 *Nature Communications* **6** 6743
[15] Pereira D *et al.* 2012 *Phys. Lett. B* **710** 426
[16] Carbone D *et al.* 2014 *Phys. Rev. C* **90** 064621
[17] Carbone D *et al.* 2017 *Phys. Rev. C* **95** 034603
[18] Cavallaro M *et al.* 2013 *Phys. Rev. C* **88** 054601
[19] Soukeras V *et al.* 2015 *Phys. Rev. C* **91** 05701
[20] Cavallaro M *et al.* 2012 *Eur. Phys. J. A* **48**:59
[21] https://web.infn.it/NUMEN
[22] Calabretta L *et al.* 2017 *Modern Physics Letters A* **32** 17
[23] Carbone D *et al.* 2016 *Results in Physics* **6** 863
[24] Muoio A *et al.* 2016 *EPJ Web of Conferences* **117** 10006
[25] Pinna F *et al. Applied Surface Science* in press
[26] Iazzi F *et al.* 2017 *WIT Transactions on Engineering Sciences* **116** 61